# Observation of the topological Dirac fermions and surface states in superconducting BaSn$_3$

K. Huang[1,2,3], A. Y. Luo[4], C. Chen[1,5,7], G. N. Zhang[1], X. L. Liu[1], Y. W. Li[1,5], F. Wu[1], S. T. Cui[6], Z. Sun[6], Chris Jozwiak[7], Aaron Bostwick[7], Eli Rotenberg[7], H. F. Yang[1], L. X. Yang[8], G. Xu[4*], Y. F. Guo[1*], Z. K. Liu[1,5*] and Y. L. Chen[1,5,9*]

[1]*School of Physical Science and Technology, ShanghaiTech University, and CAS-Shanghai Science Research Center, Shanghai 200031, P. R. China*

[2]*Shanghai Institute of Optics and Fine Mechanics, Chinese Academy of Sciences, Shanghai 201800, China*

[3]*University of Chinese Academy of Sciences, Beijing 100049, China*

[4]*Wuhan National High Magnetic Field Center and School of Physics, Huazhong University of Science and Technology, Wuhan 430074, People's Republic of China*

[5]*ShanghaiTech Laboratory for Topological Physics, ShanghaiTech University, Shanghai 201210, China*

[6]*National Synchrotron Radiation Laboratory, University of Science and Technology of China, Hefei 230029, China*

[7]*Advanced Light Source, Lawrence Berkeley National Laboratory, Berkeley, California 94720, USA*

[8]*State Key Laboratory of Low Dimensional Quantum Physics and Department of Physics, Tsinghua University, Beijing 100084, China*

[9]*Department of Physics, University of Oxford, Oxford OX1 3PU, United Kingdom*

*Email address: gangxu@hust.edu.cn, guoyf@shanghaitech.edu.cn, liuzhk@shanghaitech.edu.cn, yulin.chen@physics.ox.ac.uk*

**The interplay between topological electronic structure and superconductivity has attracted tremendous research interests recently as they could induce topological superconductivity (TSCs) which may be used to realize topological qubits for quantum computation. Among various TSC candidates, superconducting BaSn$_3$ ($T_c \sim 4.4$ K) has been predicted to be a topological Dirac semimetal (TDS) hosting two pairs of Dirac points along the $\Gamma - A$ direction. Here, by combining the use of angle-resolved photoemission spectroscopy and *ab initio* calculations, we identified the predicted topological Dirac fermions and confirmed the TDS nature of the compound. In addition, we observed surface states connecting the Dirac points. Our observations demonstrate BaSn$_3$ as a superconductor with nontrivial topological electronic structures.**

**INTRODUCTION**

Superconducting and topological quantum material are two central topics in condensed matter physics recently with intriguing physical properties and application potentials [1–4]. Their combination may further lead to a unique quantum state: the topological superconductor (TSC). A TSC can host exotic emergent particles such as the Majorana fermion which shows remarkable non-abelian statistics and serves as a key ingredient for the realization of topological quantum computation – a promising approach to realize the fault-tolerant quantum computation [5–9].

The search for TSC candidates is a challenging task. Currently, TSC candidates are only found in limited material systems, including either the superconductors with intrinsic p-wave pairing (e.g. $Sr_2RuO_4$[10, 11], $UTe_2$[12, 13] and $Cu_xBi_2Se_3$[14–16]), or artificial heterostructures combining conventional superconductors and materials with nondegenerate spin states (e.g. topological insulators[17–19], semiconductors with strong spin-orbit coupling[20, 21] or ferromagnetic atomic chains[22–24]) or compounds intrinsically possessing both the s-wave superconductivity and topological surface states (e.g., $FeTe_xSe_{1-x}$[4, 25, 26], Li (Fe, Co) As[27], $TaSe_3$[28] and 2M-$WS_2$[29, 30]).

In addition to the strong topological insulating states, the TSC and related phenomena could also be achieved with other types of topological electronic structure, such as the chiral edge states in quantum anomalous Hall insulators [31-33], surface states in topological Dirac/Weyl semimetals (TDS/TWS) [34–42] and the corner/hinge states in high order topological insulators [43, 44]. For example, the Dirac points and surface Fermi loops in TDSs would manifest as bulk point nodes and surface Majorana fermions in the superconducting states [45], respectively. With structural distortion and reduction of crystal symmetry, the nodes disappear, and a full-gapped symmetry-protected TSC is realized. The Fermi loop induced Majorana fermions are clearly distinguished from those in other TSCs [45]. In addition, topologically protected gapless bound states are predicted to exist in the superconducting TDS/TWS vortices [46]. Therefore, it would be desirable to search for TDS/TWS with intrinsic superconductivity and investigate the interplay between the superconductivity and the topological electronic states.

Currently, the combination of superconductivity and TDS states have been observed in TDS $Cd_3As_2$

(with point contact, metal junction or external pressure [47–49]), iron based superconductors Li(Fe,Co)As and Fe(Se,Te) [50], and type-II TDS $PdTe_2$ [51, 52]. However, in these materials, either the superconductivity is not intrinsic ($Cd_3As_2$), or the compound is non-stochiometric (Li(Fe,Co)As and Fe(Se, Te)), or the surface Fermi loop predicted is not on the natural cleavage termination ($Cd_3As_2$, $PdTe_2$), making the investigation of TDS-based TSC difficult. Therefore, the search for an ideal TDS with intrinsic superconductivity and topological electronic states is essential.

In this work, we report the electronic structure of a new superconducting TDS, $BaSn_3$ ($T_c \sim 4.4$ K) by presenting a systematic study of the band structure using angle-resolved photoemission spectroscopy (ARPES), with the results consistent with our *ab initio* calculation. Especially, we observed pairs of topological Dirac fermions along the $\Gamma - A$ direction and identified two surface states passing through the Dirac points, proving its TDS nature. The relatively high $T_c$, clear signature of the Dirac fermions and surface states make $BaSn_3$ a fascinating playground to investigate the elusive TSC in compounds with multiple topological electronic states.

**RESULTS AND DISCUSSIONS**

High-quality $BaSn_3$ single crystals were synthesized using the self-flux method. The crystal structure is the hexagonal lattice with space group $P6_3/mmc$ (No.194) [see Fig. 1(a) for the conventional cell]. Sn atoms form a chain structure along c direction and stack together along a and b directions to form a hexagonal lattice. The corresponding Brillouin zone (BZ) and the high-symmetry points are illustrated in Fig. 1(b), while the red box represents the (100) cleavage plane. The samples typically appear in long and narrow stripes as shown in Fig. 1(c)(i), and the high quality could be verified by the x-ray diffraction (XRD) [Fig. 1(c)(ii-iv)], from which the conventional lattice constants can be extracted as a = b = 7.254 Å, c = 5.5 Å, consistent with the previous reports [53, 54]. The superconductivity of the compound is characterized by the temperature-dependent resistivity measurement, which suggests a superconducting transition occurring at $T_c^{onset} \sim 4.4$ K, in line with the previous report [53].

The topological electronic structure of $BaSn_3$ is predicted by the *ab initio* calculations [Fig. 1(e)].

As we focus on the calculated bands near the $E_F$ along the Γ-A high symmetry lines, we find the electron-like band near Γ point with $\Gamma_8$ irreducible representation (IR) shows a linear dispersion and crosses with two hole-like bands with $\Gamma_7$ and $\Gamma_9$ IRs, respectively [shown in Fig. 1(f)]. Because of the different IRs of relevant bands, two unavoidable band crossings occur which are robust against spin-orbit interaction. Due to the protection by both time-reversal and space-inversion symmetries, each band in $BaSn_3$ is doubly degenerate, thus making the two crossings Dirac points [labeled as DP1, DP2 in Fig. 1(f)].

The general electronic structure of (100) cleavage plane is illustrated in Fig. 2. From the dispersions along the high-symmetry A-Γ-A directions measured with both the linear horizontal (LH) and the linear vertical (LV) polarized light [Fig. 2(b)(c)], we could observe the predicted hole-like bands [labeled as $\Gamma_7$, $\Gamma_9$ in Fig. 1(f)] and the electron-like bands near Γ point [labeled as $\Gamma_8$ in Fig. 1(f)]. Besides, several other bands are observed, in line with the slab band calculation [Fig. 2(d)]. The constant energy contours (CECs) at different binding energies [Fig. 2(a)] demonstrate the evolution of the band structure in the 2D BZ, which proves the hole/electron nature of the $\Gamma_7$, $\Gamma_9/\Gamma_8$ bands and shows nice agreement with the theoretical calculations [Fig. 2(a)].

To investigate the evolution of the electronic structure along the $k_z$ direction, and also the surface bulk origin of the observed bands, photon energy-dependent experiment was performed with the result shown in Fig 3. The CEC at binding energy 1.5 eV in the $k_z$ - $k_y$ plane is plotted in Fig. 3(a), where clear periodic patterns were observed with a periodicity of $2\pi/c$ [the BZ is indicated by the red dashed lines with high-symmetry points labeled], allowing us to determine the high symmetry points in the 3D BZ [such periodicity could also be observed from the cut along the A-L-A direction extracted from the photon energy dependent measurement, see Supplemental Material at [55] for Detailed photon energy dependence data]. The systematic measurement of the dispersion from the Γ-A-Γ to M-L-M direction allows us to visualize the band evolution near the predicted Dirac point.

Fig. 3b(i) shows the band dispersions along the A-Γ-A direction measured with 92 eV photons (corresponding to $k_z\sim 2n\pi/c$). By tracking the peaks according to the $\Gamma_8$ and $\Gamma_9$ bands [indicated as α and β bands in Fig. 3b(i)] in the stacked momentum distribution curves (MDCs) plot [as indicated by the purple dashed line in Fig. 3b(ii)], we could find both of them have linear dispersions and cross with each other at

~0.7 eV below $E_F$ [labeled as DP in Fig. 3b(ii)]. We could not resolve the theoretically predicted two adjacent Dirac points (DP1 and DP2) due to their adjacency which go beyond the resolution of our instrument and that the $\Gamma_7$ mixes with $\Gamma_9$. The dispersion parallel to the A-$\Gamma$-A direction with different $k_z$ values are illustrated in Fig. 3(c)-(f) [their positions in the BZ is labeled in Fig. 3(a)]. Evidently, for increased $k_z$ value, there is a gradual spectral weight transfer from $\alpha$ band to $\alpha$' and $\gamma$ bands while the $\beta$ band loses its spectral weight, leading to the opening of a gap at the DP [the gap is clearly visible in the stacked MDC plot at $k_z$=0.8$\pi$/c, see Fig. 3(f)(ii)]. Such $k_z$ evolution shows nice agreement with the theoretical calculations [see Supplemental Material at [55] for detailed calculations of the bulk band structure at different $k_z$ value], confirms that DP is the Type-I Dirac point and $\Gamma_8$ and $\Gamma_9$ bands form bulk Dirac fermions in BaSn$_3$.

In addition, from the photon energy dependent data measured with different polarization, we could distinguish the surface originated bands crossing the DP. Fig. 4(a)-(i) shows the comparison of the dispersion parallel to A-$\Gamma$-A taken at different $k_z$ value with both linearly horizontal (LH, right)/linearly vertical (LV, left) polarized photons [see Supplemental Material at [55] for the measurement geometry]. While the bulk bands show the gap-opening behavior near the DP as suggested by the measurement using LH polarized photons [Fig. 3(b)-(f) and right half of Fig. 4(a)-(i)], we could observe one feature crossing DP showing absence of $k_z$ dependence [labelled as SS1 in Fig. 4(a)-(i) and the CEC in the $k_z$-$k_y$ space at 0.65 eV in Fig. 4(k)(i)]. Due to the photoemission matrix element effect, bands with different orbital components could be enhanced/suppressed. In the band structure measured with LV polarized photons [see the left half of Fig. 4(a)-(i)], another feature manifests itself [labelled as SS2, see Fig. 4(k)(ii) and also labelled on the CEC in the $k_z$-$k_y$ space at 0.81 eV]. The absence of $k_z$ dispersions of SS1 and SS2 suggests their surface origin. As both SS1 and SS2 pass through DP, it possible suggests that they might be the "surface Fermi loop", or topological surface states as previously reported in Ref. [45]. These surface states have been predicted to exist in the calculated spectrum for semi-infinite system along the $\overline{A}$ - $\overline{\Gamma}$ - $\overline{A}$ direction [as shown in Fig. 4 (j)], but their topological origin could not be easily identified due to the intertwined nature of the surface-bulk hybridization.

The complicated surface-bulk hybridization has also been observed in type-II TDS family [56].

However, the nontrivial helical spin texture preserves even in the topological surface resonance states [56], showing its robustness against surface-bulk hybridization [57]. Therefore, as a prerequisite for effective *p*-wave superconductivity [58], the helical spin texture of the surface states requires further investigation.

Furthermore, as the coexistence of the Dirac points and the "surface Fermi loop" is ensured by the topological invariant [45] and the mirror-reflection symmetry in $BaSn_3$, the mirror Chern number of the superconducting Dirac semimetal is nonzero, resulting in double Majorana fermions on the mirror-invariant line [45], if the gap function of this compound is mirror-odd. And the mirror-odd gap function of the TSC is detectable via anomalous Josephson effects [59] with carefully fabricated junctions. Therefore, our ARPES measurements reveal the presence of the topological electronic state and could encourage the theoretical and experimental researchers for an in-depth examination of superconductivity in $BaSn_3$.

**CONCLUSION**

In conclusion, our ARPES measurement and *ab initio* calculation reveals the complete electronic structure of $BaSn_3$. The excellent agreement between the experiment result and the calculation strongly supports the superconducting $BaSn_3$ as a TDS with the topological Dirac fermions along the A-Γ-A high-symmetry direction and the existence of two surface states passing through the Dirac points. Our results demonstrate $BaSn_3$ as a promising TSC candidate not only for the study of exotic properties of multiple topological electronic states in a single material, but also the interplay between nontrivial topological states and superconductivity.

**METHOD**

**Sample synthesis**

High quality $BaSn_3$ crystals were grown from a self-flux method. The Ba pieces and Sn lumps were mixed in the molar ratio 1: 6 and sealed in an evacuated quartz tube under the vacuum of about $10^{-4}$ Pa. The tube was slowly heated to 700 ℃ in 15 h, kept at this temperature for 10 h in order to mix the solution uniformly, and then slowly cooled down to 400 ℃ at a rate of 3 ℃ /h. The single crystals were obtained by immediately putting the tube in a high speed centrifugation at this temperature to remove the excess Sn. Columnar hexagonal shaped crystals were obtained after breaking the tube in the glove box protected by high-purity Argon gas.

*Ab initio* **calculation**

The present calculations have been performed using DFT code VASP [60–62], which is an implementation of the projector augmented wave (PAW) method. Perdew-Burke-Ernzerhof (PBE) type of the generalized gradient approximation [63] was used as the exchange-correlation potential. The experimental lattice constants (a = b = 7.254 Å, c = 5.5 Å) are taken, and inner positions are obtained through full relaxation with a total energy tolerance $10^{-6}$ eV. The cut-off energy for the wave function expansion was set to 400 eV, and a $7 \times 7 \times 9$ k-mesh in the first Brillouin zone (BZ) was used for self-consistent calculations. The SOC was considered self-consistently. We then construct the maximally localized Wannier functions for the 5p-orbitals of Sn using WANNIER90 [64] and calculate the topological properties using WannierTools [65].

**Acknowledgements**

We wish to thank Prof. G. Li and Y. Y. Y. Xia for fruitful discussions. This research used resources of the Advanced Light Source, a U.S. DOE Office of Science User Facility under contract no. DE-AC02-05CH11231, and other two ARPES beamlines: BL03U of Shanghai Synchrotron Radiation Facility and BL13U of National Synchrotron Radiation Laboratory. We also acknowledge the Analytical Instrumentation Center of ShanghaiTech University for X-ray and Laue diffraction measurements. This work was sponsored by the National Key R&D Program of China (grant No. 2017YFA0305400 to Z.K.L. and grant No.2018YFA0307000 to G.X.), the National Natural Science Foundation of China (grant No. 11674229 to Z.K.L., grant No. 11874022 to G. X. and grant No. 12004248 to H.F.Y.), the Major Research Plan of the National Natural Science Foundation of China (No. 92065201 to Y. F. G.), Shanghai Municipal Science and Technology Major Project (grant No. 2018SHZDZX02 to Y.L.C. and Z.K.L.), Shanghai Sailing Program (grant No. 20YF1430500 to H.F.Y.).

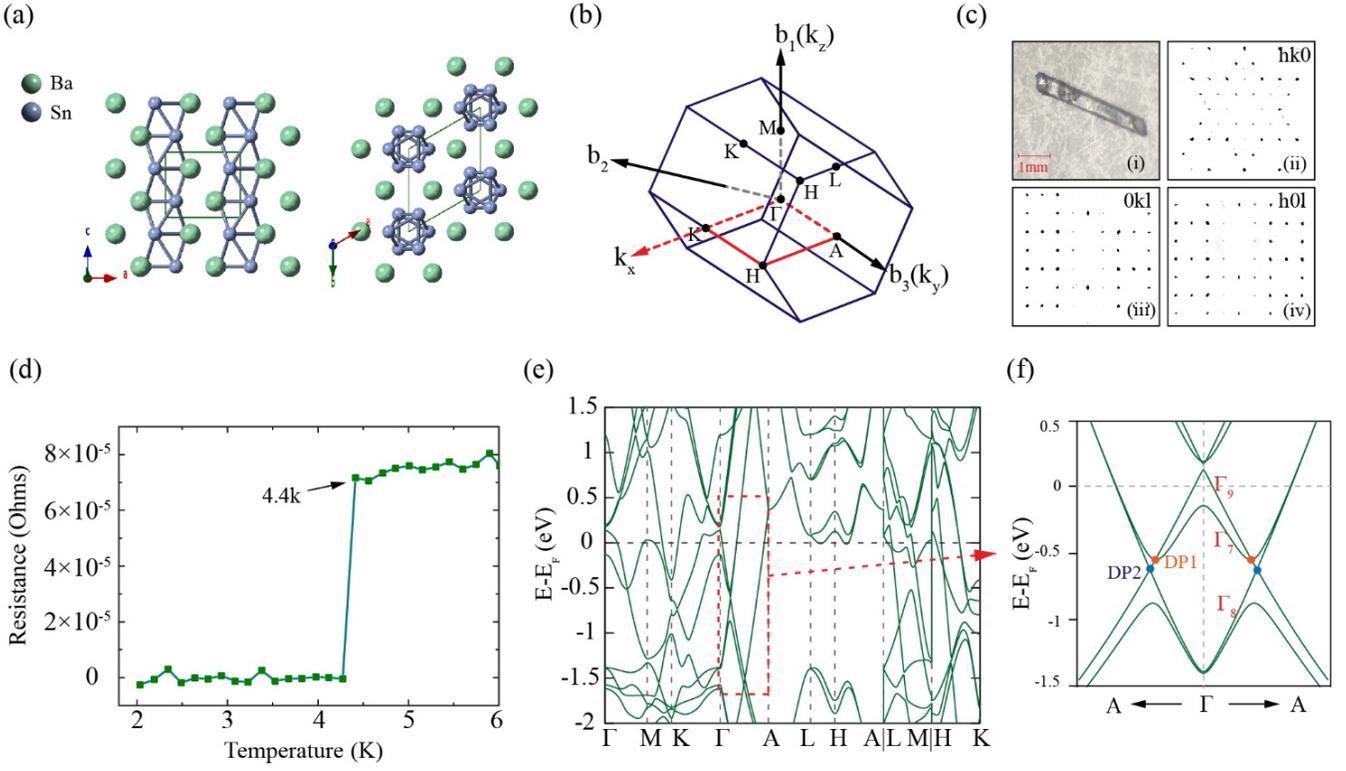

FIG. 1. Basic characterization of BaSn₃. (a) Crystal structure of BaSn₃, green and blue spheres indicate Ba and Sn ions, respectively. (b) The corresponding BZ of BaSn₃; high-symmetry points are labeled. Red box indicates the (100) cleavage plane. (c) Image of BaSn₃ single crystal (i) and X-ray diffraction patterns along three different directions (ii-iv). (d) The temperature dependent resistivity of BaSn₃ and the superconductivity transition onset temperature is marked. (e) Calculated bulk electronic structure of BaSn₃ with spin-orbit coupling. (f) Enlarged view of the bulk band dispersion along the A-Γ-A direction and labeled IRs of the bands forming DPs.

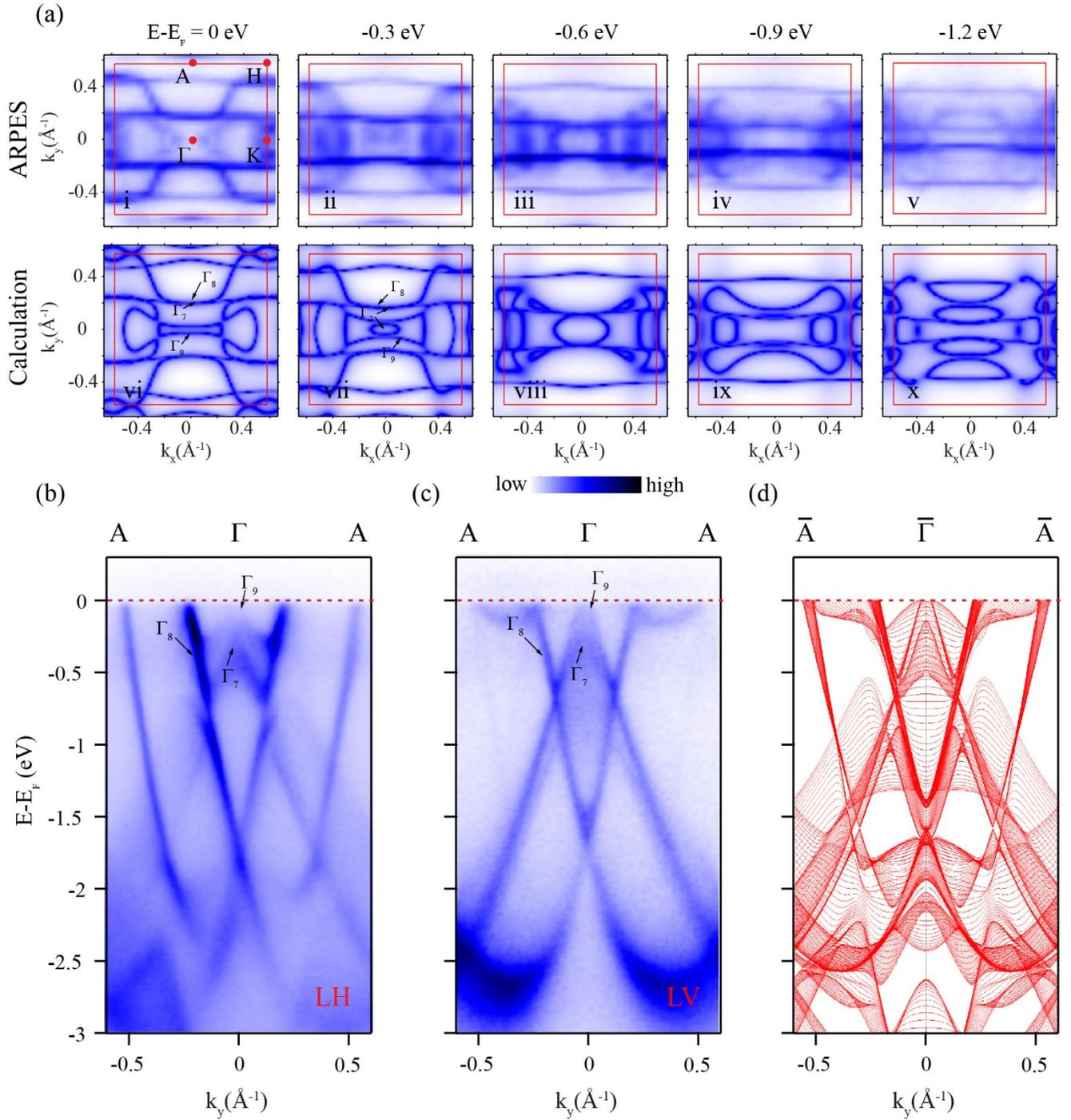

FIG. 2. Basic electronic band structure of BaSn$_3$ of the (100) surface. (a) (i-v) Photoemission intensity map of constant energy contours (CECs) in the k$_x$-k$_y$ plane at 0, 0.3, 0.9, 1.2 eV below E$_F$, respectively. (vi-x) Corresponding theoretical ab initio calculation of CECs at the same energy. (b)(c) Band dispersions along high-symmetry A-Γ-A measured with LH and LV polarized photons. (d) Slab band calculation of dispersions along the high-symmetry $\overline{A}$ - $\overline{\Gamma}$ - $\overline{A}$ direction. CECs in (a) are symmetrized according to the crystal symmetry. The integrated energy window is ±50 meV. The data were collected using photons with hν = 92 eV (corresponding to k$_z$ ∼ 0) at 20 K.

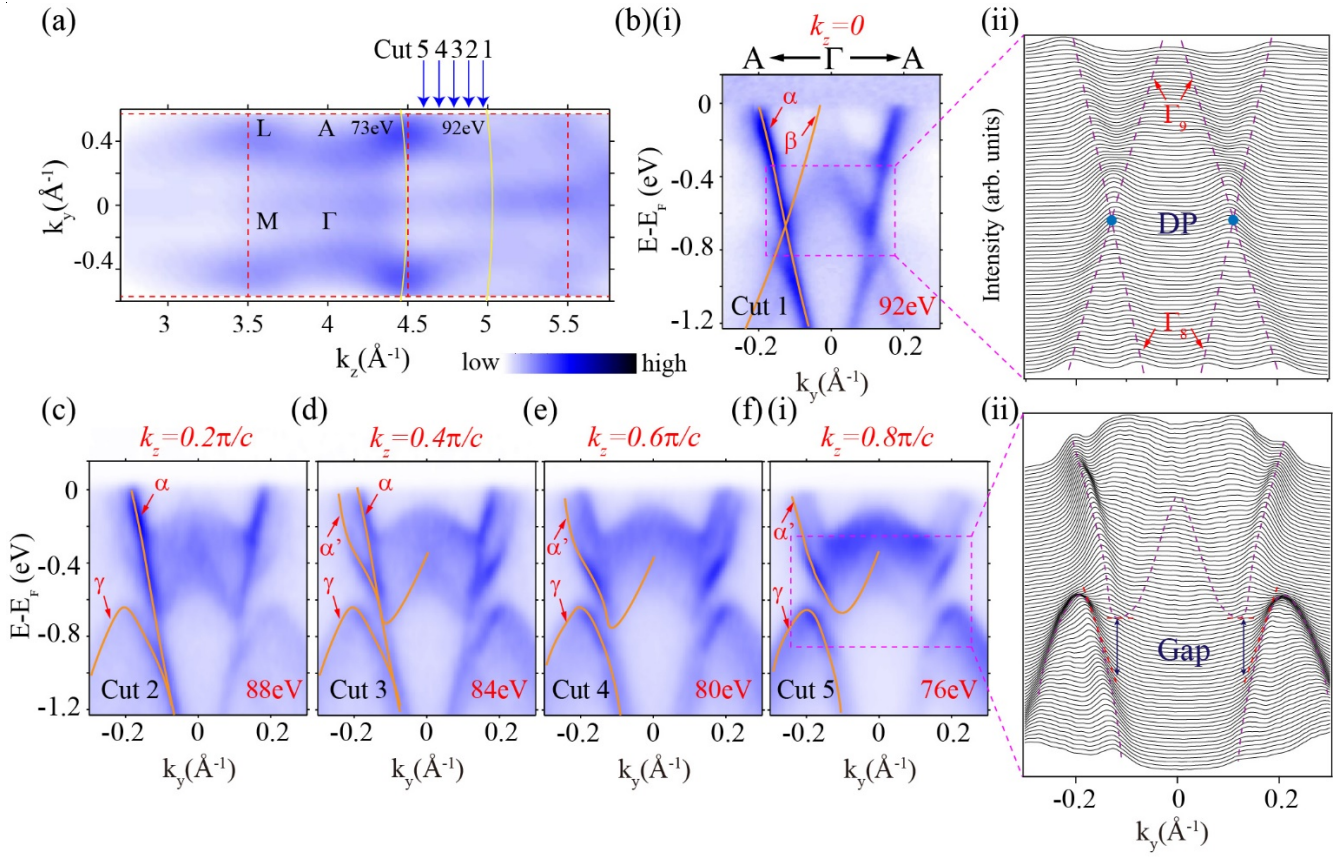

FIG. 3. $k_z$ evolution and the evolution of the DP in BaSn$_3$. (a) Photoemission intensity map of the CEC in the $k_z$-$k_y$ plane at 1.5 eV below $E_F$. Red dash line indicates the BZ in the (110) direction. Arrows indicate the cut directions shown in (b)-(f). (b)-(f) Band dispersions along high-symmetry A-$\Gamma$-A directions with different photon energy, respectively. The bands discussed in the manuscript are labeled. b(ii)f(ii) Zoomed-in stacking plot of momentum distribution curves (MDCs) of b(i) and f(i). The data were collected with LH polarization.

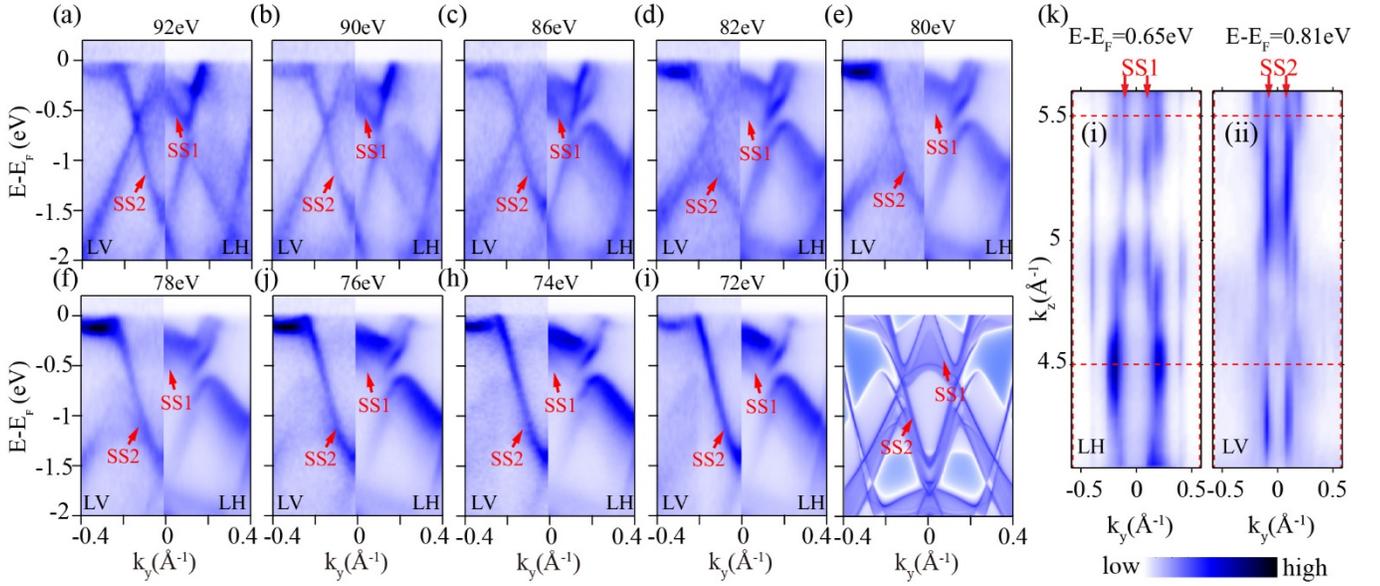

FIG. 4. Observation of the surface states crossing the DP in BaSn$_3$. (a)-(i) Band dispersions along high-symmetry A-Γ-A with two polarization and different photon energy. The left part shows the band dispersions measured with LV polarized photons while the right part shows the band dispersions measured with LH polarized photons. SS1 and SS2 label the surface states identified. (j) Surface state spectrum for semi-infinite system calculation along high-symmetry $\overline{A}$ - $\overline{\Gamma}$ - $\overline{A}$ direction. (k) Photoemission intensity map of the CEC in the k$_z$-k$_y$ plane at 0.65 and 0.81 eV below E$_F$, respectively. The CEC in (i) is measured with LH polarized photons while in (ii) is measured with LV polarized photons. Red dash line indicates the BZ of the (110) surface.

405 (2018).